\documentclass{aa}
\usepackage{epsf,psfig}

\begin{document}
\thesaurus{ 02(02.07.1; 03.13.4; 12.03.4; 12.12.1) }        

\title{Non-linear dynamics and mass  function of cosmic structures. II
  Numerical results}
\author{Edouard Audit$^1$, Romain Teyssier$^2$
  and   Jean-Michel     Alimi$^1$}     \institute{$^1$     Laboratoire
  d'Astrophysique  Extragalactique et de  Cosmologie, Observatoire  de
  Paris-Meudon, F-92195 Meudon Cedex,
  France\\
  $^2$  CEA,  DSM/DAPNIA/Service d'Astrophysique  CE-SACLAY,   F-91191
  Gif--sur--Yvette,   France}  
\date{} 
\maketitle 
\markboth{Non-linear-Dynamics and the mass function}{II Numerical Results}
% my definitions
\def\etal{{\it et al. \/}}
\def\eg{{e.g.,\ }}
\def\etc{{etc.\ }}
\def\ie{{i.e.,\ }}
\unitlength=1cm

\begin{abstract}
  We  compare  the  mass  functions   obtained  analytically, in   the
  framework of  an extended Press \& Schechter  (PS)  formalism, in a
  previous paper to the numerical mass functions obtained in N -- body
  simulations,  using different algorithms  to  define objects in  the
  density  field.  After discussing  the properties of the algorithms,
  we show  that the mass  function obtained using  the friend -- of --
  friend algorithm reproduces best the  scaling behaviors predicted in
  the extended PS formalism.  Following this statistical analysis,  we
  show  that it is   possible  in the   framework of  our  extended PS
  formalism to reproduce the mass function but also, and for the first
  time, the  initial statistical  properties  of structures  and their
  collapse  time. This   allow to present   a  ``coherent'' picture of
  structure  formation which can   account for the  initial, final and
  dynamical properties of structures.
\end{abstract}

\section{INTRODUCTION.}

This   paper is  the  second  part  of a   study  on  the influence of
non--linear dynamics on the  mass function of  cosmic structures. In a
companion paper (\cite{AA97}, thereafter  P1) we  studied analytically
the influence of the shear  and of the tide  on the mass function in a
generalized Press \& Schechter  (1974, thereafter PS) formalism for  a
critical  universe.  We first  showed that  the  mass function can  be
directly   related to a  given  dynamical   model through a  selection
function  which gives the probability  that a Lagrangian fluid element
with a given initial density contrast will be in a collapsed structure
at the present epoch.  We also emphasized two invariance properties of
the mass function in  such a formalism, the well-known self-similarity
in time, and  the possibility of  factorizing the mass function as the
product of  a function which depends only  on  the dynamical model and
another function  which depends only on the  power spectrum of initial
density fluctuations (P1).  These time and spectral scaling properties
are   strong predictions of the   generalized  PS formalism, they will
therefore be used to test the validity of such an approach.

In the present paper, we  extend our analytical  work by a comparative
analysis with numerical simulations. These ones have been used by many
authors to  test the  PS  approach.  The first  successful attempt was
made  by Efstathiou et  al.  (1988)  using  however a small number  of
particles.  They  used a rather simple  and easily tractable algorithm
to define collapsed  objects in the  numerical density  fields, namely
the Friend -- of -- friend algorithm  (thereafter FOF).  They conclude
that the standard PS mass function reproduces well the numerical data,
although  the number of  objects with mass $M \simeq  M_*$ seems to be
overestimated  ($M_*$ precedes the  cut-off  at large mass).   Gelb \&
Bertschinger (1994) have  presented  a  new algorithm (DENMAX)   which
brings together  particles around maxima of   the density field.  They
showed  that  the resulting mass function   differs from the  FOF mass
function.     Lacey \& Cole  (1994)  compared  also the mass functions
obtained with different algorithms.  They used  mainly FOF and another
algorithm that detects  spherically symmetric  density maxima in   the
field (spherical  overdensity, thereafter  SO).  These works emphasize
that   the resulting  mass  function   depend strongly   on the chosen
algorithm to    define structures in  the  density  field.   A similar
conclusion was  outlined  by Eke  et al.  (1996)   in the case  of the
correlation function of clusters in a CDM cosmogony.

In this paper we  work then towards two  goals.  First, we study as in
the previous  works,    the dependence on    different  definitions of
structures, of  the mass functions  in  the numerical simulations.  In
particular,   we examine if  the  resulting mass functions satisfy the
time and spectral scaling properties,  inherent to a PS description of
structure formation.  Secondly,  we investigate the connection between
the actual fully non  -- linear dynamics of   structures in N --  body
simulations and the dynamical model used  to compute the mass function
in analytical approaches.  As a matter of  fact, even if the number of
objects in the field can be relatively well described by, for example,
the  standard  PS mass function,  as   was argued in  previous studies
(\cite{EFWD88},  \cite{LC93}),  this however  does   not mean that the
actual dynamics of each object is correctly described by the spherical
model.  Finally, we build a complete  description of the mass function
which  allows both a correct statistical  description of the number of
objects generated    in the numerical  simulations,  but  also a right
description of  their  initial statistical  properties   and of  their
dynamics.
  
In the next section we present our numerical simulations of scale free
power spectra  and the different  algorithms we used to define objects
in the density field. The resulting mass functions and the analysis of
their scaling    properties are discussed   in  sect.  3.   A complete
description, which accounts   for both the statistical  properties and
dynamical  origin of the  mass function is exhibited  in the sect.  4. 
Finally we discuss our results and conclude in sect. 5.

\section{NUMERICAL SIMULATIONS}

We consider  three power-law initial    power spectrum ($P(k)  \propto
k^{n}$ with  $n=0,-1,-2$)  in  a  critical  universe with   a   Hubble
parameter $h=0.5$.  The simulations were done using a Particle -- Mesh
code with   $128^3$ particles and  $256^3$ grid  points to compute the
gravitational force.  The interpolation between the particules and the
grid points was  done with a  Cloud-In-Cell interpolation scheme.  The
initial power  spectrum    were  all normalized with    the  condition
$\Delta(R_*=8h^{-1}Mpc)=1$,  where $\Delta$    is  the  r.m.s  density
fluctuation  filtered by a Top-Hat  window function of radius $R$, and
linearly extrapolated to present time.   For scale free initial  power
spectra, $\Delta$ and the mass $M$ are related by

\begin{equation}
\Delta(M) = \left( \frac{M}{M_*} \right) ^{-\frac{n+3}{6}}
\end{equation}

\noindent where $M_*=(4\pi /3)\rho_0   R_{*}^{3}$  defines  a  
characteristic mass which marks  the transition between linear and non
-- linear scales ($\Delta (M_*) = 1$). Any numerical simulations which
intend to test  the PS approach of  structure formation should be able
to describe correctly  scales where $\Delta (M) \ll  1$ down to scales
where $\Delta (M) \gg 1$.   Consequently, the choice of the simulation
box size is of  crucial importance.  This choice  is then a compromise
between  two requirements:  the  resolution to  describe correctly the
dynamics of a single cluster and the  power on large scales.  This has
lead us to choose a box-length of $L=80h^{-1}Mpc$ for $n=0$ and $n=-1$
and  of $L=200h^{-1}Mpc$ for  $n=-2$ which  has  more  power on  large
scales.  This    gives,  with the   above  normalization, $M^{*}=8700$
particles for $n=0$ and $n=-1$ and $M^{*}=560$ particles for $n=-2$.

In order  to consider  only the structures  for  which the dynamics is
properly described by the PM code,  we have kept only structures whose
virial radius (defined as the radius of the sphere  for which the mean
overdensity is $180$) are greater than $1.5$ cells which is the radius
at which the numerical force equals only half of the true force.  This
conditions gives  a minimum mass of   $M_{min} \simeq 300$  particles. 
This   limit  is very  conservative  but  it  ensures  that there  are
negligible resolution effects.  For each spectrum we have done several
simulations ($3$  for $n=0$,  $4$ for $n=-1$  and  $5$ for  $n=-2$) to
increase the statistical significance  of our results.  We obtain, for
each spectrum, a few thousand  well described ($M>M_{min}$) objects at
an expansion factor $a=1$ (present time).   All the results we give in
this paper are statistical averages over all the simulations performed
for each power spectrum.

The  time  variable was  chosen in order  to  obtain  a constant r.m.s
displacement   of  particles,   independently  of  the   initial power
spectrum.  This can be achieve with the time variable $p = a^{\alpha}$
where $\alpha = 2/(n+3)$ (Efstathiou et al. 1985). Each simulation was
run with the same number of time steps ($500$) (using a leap-frog time
integrator), which ensures an excellent energy conservation (see Table
[1]).

\begin{table*}
\begin{center}
\begin{tabular}{|c|c|c|c|c|}
\hline
n & \# of & $L_{box}$      &  outputs times    & average    \\
  & runs  & (Mpc h$^{-1}$) &(expansion factors)&$\Delta E$ (\%) \\
\hline 
\hline
0  &       3 & 80 & 0.18 0.25 0.35 0.50 0.70 1 &   0.8 \\
\hline       
-1 &       4 & 80 & 0.31 0.40 0.50 0.63 0.79 1 &   0.9 \\
\hline
-2 &       5 & 200& 0.56 0.63 0.70 0.79 0.89 1 &   0.7 \\
\hline
\end{tabular}
\caption{Parameters of PM simulations used in this paper. The grid
  size is $256^3$,  and  the number of particles  is  $128^3$ for each
  run.   $\Delta E$ is  the energy  conservation (in  percent) for the
  entire simulation averaged over all the runs for a given spectra.}
\end{center}
\end{table*}

In order  to study the  time evolution of the  mass function we stored
the results  of the simulations  at $6$ different epochs.  Each output
corresponds  to $M_{*}$ equal  to twice the last  output value.  Table
[1] gives the different simulations parameters we used for this study.

\subsection{Definition of objects in the density field}

To count the   number    of structures generated  in   a  cosmological
numerical  simulation, one  needs    to  specify   what we  mean    by
``structure''.  We  have  then   analyzed the   three  main  type   of
algorithms that   define  differently  collapsed objects in    a given
density  field, respectively Friend  -- of -- friend (thereafter FOF),
DENMAX (\cite{GB94}),   a  modified  version of DENMAX   that  we call
DENEVAP, and  Spherical  Overdensity.  The quality  and drawbacks that
one   generally  imputes to these   algorithms  cannot be justified by
rigorous physical arguments.    In  our work we  favor   an algorithm,
because it defines  structures, in such a way  that the resulting mass
function satisfies the time and spectral scaling invariances, inherent
to the PS formalism.

\subsubsection{Friend -- of -- Friend Algorithm}

This algorithm is  one of the most  used methods to define objects. It
is based on a percolation scheme that links particles recursively when
the spatial  separation between  two particles  is less than   a given
threshold,   called   the percolation length   $l$.    This  length is
expressed in units of  the  initial mean interparticular separation.   
This method  is equivalent  to selecting regions   of space  which are
enclosed    by a density  isocontour  of  $1/2l^3$.  For a spherically
symmetric object   with an isothermal  density  profile  $\rho \propto
r^{-2}$, this  leads to a  mean interior  density $\bar{\rho}/\rho_c =
3/2l^3$.  The  choice $l=0.2$ gives  a mean density contrast of $187$,
which is the value  generally assumed for the  final relaxed object in
the  framework of  spherical    collapse.  We don't however   restrict
ourselves to this  value,  as  we will see   in the  following.    The
advantages of this algorithm are its easy  implementation and the fact
that there is only one free parameter which can be related directly to
a   density   isocontour.  Moreover,  this    scheme  doesn't make any
assumption about the actual shape of objects.  The main drawback is an
overlapping  problem    (\cite{BG91}, \cite{GB94}).    Indeed,   small
filaments can connect two objects which are actually different.  These
bridges   of  matter lead to a   single  massive object  which has not
collapsed, which could affect the mass function.

\subsubsection{Denmax Algorithm and Modifications}

The  last  remark has  led   Bertschinger \& Gelb   to develop   a new
algorithm (DENMAX) to define structures  in the density field.  DENMAX
is based  on a density maximum   detection, linking together particles
belonging  to the same density peak.   Particles are moved towards the
peak  position following   geodesics   of  the density   field.    The
surrounding surface  of  the corresponding object  is  therefore not a
density  isosurface, but the col  surface  delineating the ``influence
region'' of a  peak.  This scheme does not  suffer from an overlapping
problem. It has however a very important  drawback. The density maxima
and the corresponding regions greatly depend on the scale at which the
density field is  computed.    This  means  that  DENMAX has   a  free
parameter, which is the spatial resolution of the grid used to compute
the density field. This leads to resolution dependent results. One has
to be careful when choosing the grid resolution.  A second drawback of
the  method is that  it defines objects  without  an intrinsic density
criterion.  Indeed, very low  density peak are identified as  objects,
although these small density ripples  are not collapsed objects.  Gelb
\& Bertschinger (1994) used a additional criterion to remove particles
that are   not  gravitationally bound  to  the  peak  which   they are
associated with.  Here, we rather use  an overdensity criterion.  This
second  algorithm,  which is a   modification  to the original DENMAX,
evaporates  objects which do not   satisfy the following requirements:
for each  density maximum,  we apply  the FOF  algorithm with  a given
percolation length $l$, and  we extract the most massive  sub -- halo. 
We then compute its center of mass, and its three  principal axes.  We
then define the mean interior density  $\bar{\delta}$ of the object as
the total mass divided by the volume of the  mean quadratic ellipsoid. 
We vary the percolation length   until the overdensity reaches $180$.  
With this new method, low density peak are totally evaporated, and low
density wings surrounding  halos do not contribute  to the final  halo
mass.  For a three dimensional object with a  complex shape , defining
the mean interior density is a difficult task.  This criterion appears
to be a good   compromise, since we  checked  that in  the case of   a
spherical isothermal halo, we recover the value of $180$ corresponding
to  the spherical model.  We refer  to this modified version of DENMAX
as DENEVAP.   Note however that this  new method still depends  on the
resolution used to compute the density field,  and that the first step
of DENEVAP is still the DENMAX algorithm.

\subsubsection{Spherical Overdensity Algorithm}

This last method is  a tentative algorithm  described in Lacey \& Cole
(1994) that  avoids  the drawbacks of  the  two previous schemes.   It
consists of  finding spherical  regions  in  the  simulation  having a
certain mean overdensity  (for example $\bar{\delta}=180$).  We  first
calculate a local density for each particle using a nearest grid point
affectation    scheme.   We then  sort  the   particles  by density in
decreasing order.  The  center of mass  of a halo is first  identified
with the  first particle of the  list.  We grow  a sphere  centered on
this point  until the mean overdensity  reaches the chosen  value.  We
then  calculate the new center of  mass of the  particles contained in
this sphere, and we iterate until the distance between two consecutive
center of mass is less than 10\% of the mean interparticular distance.
We finally remove  all particles of the new  halo from the list.  Note
that the  halos  do not  depend on  the  method we use  to compute the
density, and to find   the density maxima, as   long as we use a  high
resolution grid  (we  use   a $1024^3$ grid    in this  paper).   This
algorithm is thus  resolution  independent.  There is no   overlapping
problem,  and  there is  only  one  free parameter, namely  the chosen
overdensity threshold.  However,   this scheme  has the  drawback   of
imposing a spherical shape on the halos.

\begin{center}
\begin{figure*}[httb]
\psfig{file=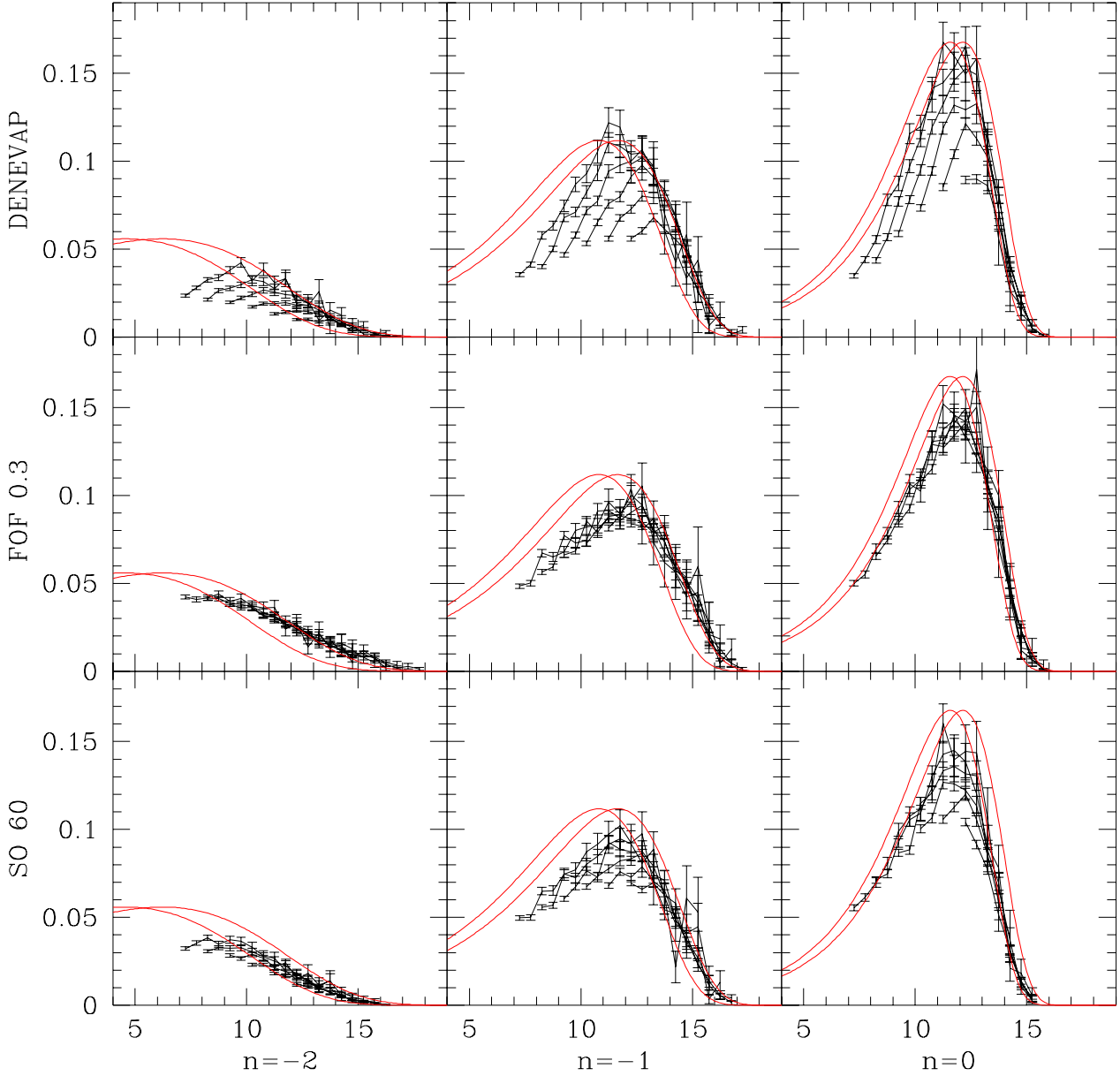,height=18.0cm}
\caption{Multiplicity functions obtained for DENEVAP, SO60 and FOF-0.3
  and for the different initial power  spectra. Each curve corresponds
  to a  different output time rescaled to  $a=1$. The two  solid lines
  represent   the   PS  mass    function with  $\delta_c=1.4$ (largest
  high-mass cut-off) and $1.686$.}
\label{fmlog1}
\end{figure*}
\end{center}

\begin{center}
\begin{figure*}[httb]
\psfig{file=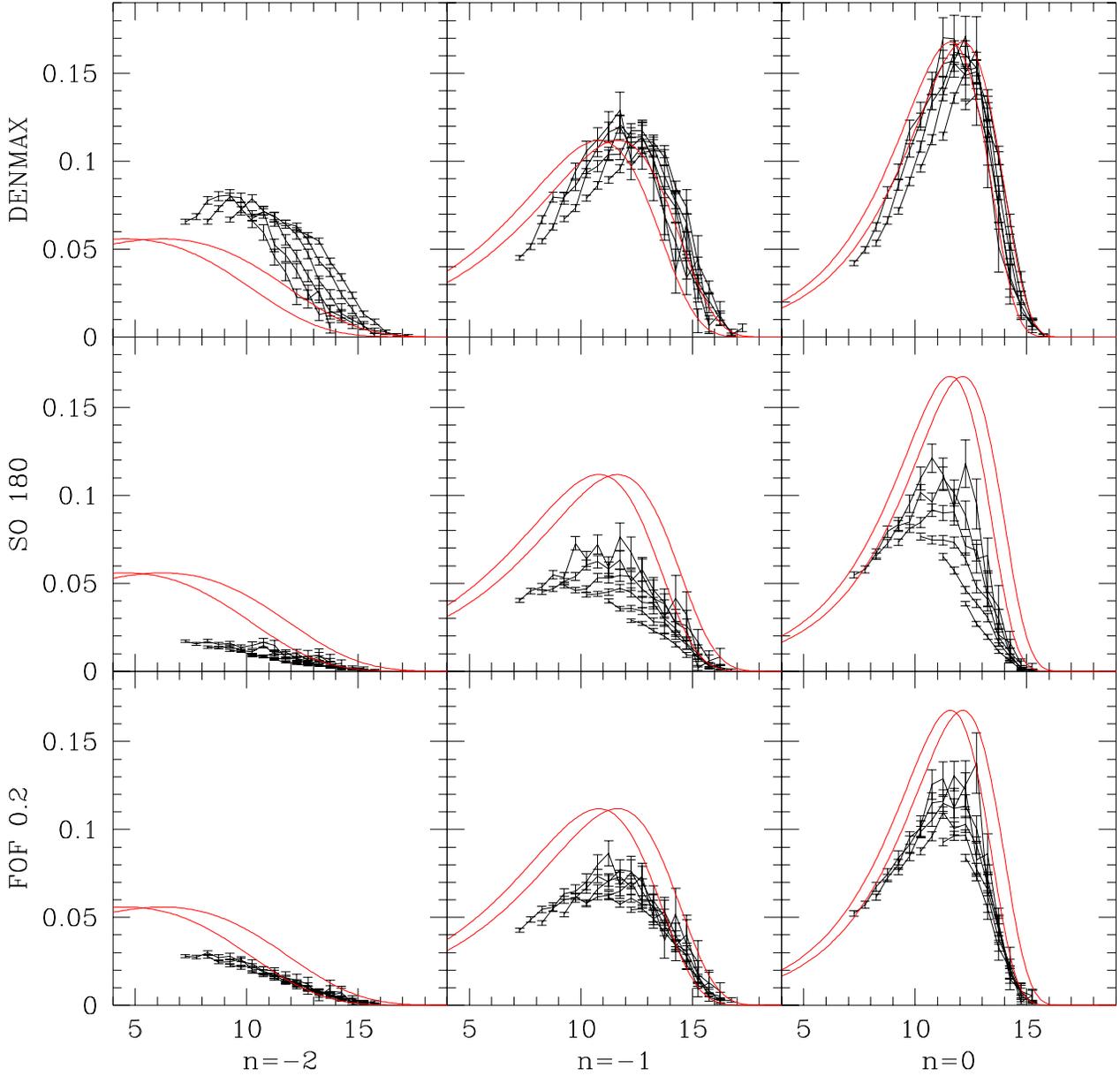,height=18.0cm}
\caption{Same as figure (\ref{fmlog1}) but for DENMAX SO180 and FOF-0.2.}
\label{fmlog2}
\end{figure*}
\end{center}

\section{Multiplicity Function}

In this  section we  present  the mass function  obtained at different
times, for  the  different power   spectra   and for the    algorithms
presented   above.  When looking at  these  mass functions we have two
goals.  First we  want to look  at their  dependence on the  different
methods  to define structures.   Then, we want to  see if they satisfy
the time self -- similarity and the spectrum  invariance that are very
generic features of  the mass function obtained through  a  PS -- like
approach (P1, see Appendix).

%Second, if they satisfy these two criteria, we want to determine which
%dynamical prescription is  favoured, if  any.  In  the second  part of
%this   section we  look  at   the  dynamical properties  of structures
%individually  to see   whether   the dynamics suggested   by  the mass
%function is indeed the one followed by actual structures.

Following Efstathiou et al.   (1988), we use the multiplicity function
defined as  the fraction of  mass embedded in  halos with a  number of
particles   between   $2^{m}$  and   $2^{m+1}$,   where $m$   is   the
multiplicity.  In these terms,  the multiplicity of the characteristic
non -- linear mass $M_*$  between two consecutive outputs increases by
one.   In Figs.   \ref{fmlog1}  and  \ref{fmlog2}  we present    the
multiplicity functions   obtained for  each   spectrum  and  for  each
algorithm.  The multiplicity  functions obtained   at each  time   are
rescaled to $a=1$, in order to test the time self -- similarity of the
mass function. Note  also that  the mass (in  number of  particles) of
halos obtained in the  $n=-2$ simulations were  multiplied by a factor
$2.5^3$ to account for the different box size.  The error bars in both
figures are computed using a Poisson noise estimation in each bin.  We
now discuss the multiplicity functions obtained with each algorithm.

\subsection{FOF}

As we have seen this algorithm has one free parameter, the percolation
length $l$.  We have first tried the value $l=0.2$, adopted by several
authors.  Our results (Fig.  2) are in  good agreement with Efstathiou
et al.  (1988).   The time self  similarity is very well recovered and
the high    mass  cut off is    well reproduced  by   the  standard PS
multiplicity function ($\delta_c=1.686$,  see appendix).  However,  as
mentioned by Efstathiou et al., there is  a mass deficit of about 50\%
around the peak  of the multiplicity  function.  This deficit suggests
that  the density criteria is too  stringent.  Therefore we have tried
FOF  with  $l=0.3$.   This   roughly corresponds  to  halos   with  an
overdensity  of $60$.  In this  case the  cut-off  corresponds to a PS
multiplicity function  with $\delta_c=1.4$  and the discrepancy around
the peak is   considerably reduced  (Fig.    1).  Since for   both the
percolation   lengths  used  the  time  self-similarity  is  very well
recovered it is possible to average the multiplicity function obtained
at different times (rescaled to   $a=1$) to get a unique  multiplicity
function   for  each  spectra.  One    can then  check  that  all  the
multiplicity  functions obtained  for  the different  spectra have the
same  universal   multiplicity function  (see    appendix  or P1   for
definition).   We have plotted on Fig. 4-A (resp.4-D) the universal
multiplicity function  for $l=0.3$  (resp.  $l=0.2$).  The  ``spectral
scaling'' predicted in the  framework  of a  PS approach is  extremely
well recovered.

For the FOF  algorithm, and for  both percolation lengths, the generic
features (time and spectral scaling) predicted by the PS formalism are
well reproduced. For a percolation length  of $0.2$ the cut-off of the
multiplicity  function  is   well reproduced   by   the  spherical  PS
multiplicity  function  but in that  case  there is a  large excess of
``analytical''  structures around $M_*$.     For a larger  percolation
length  ($l=0.3$)   the entire multiplicity    function is  quite well
reproduced by a PS function with $\delta_c=1.4$  (even though there is
still  a small excess of ``analytical''structures   around $M_*$).  We
have shown in P1 that  a PS-like function with $\delta_c=1.4$  results
from the PS  formalism with a non-spherical  dynamics.  We will see in
the  next section if  this agreement can   really be explained by this
dynamical reason.

\subsection{Denmax}
\begin{figure*}[httb]
\psfig{file=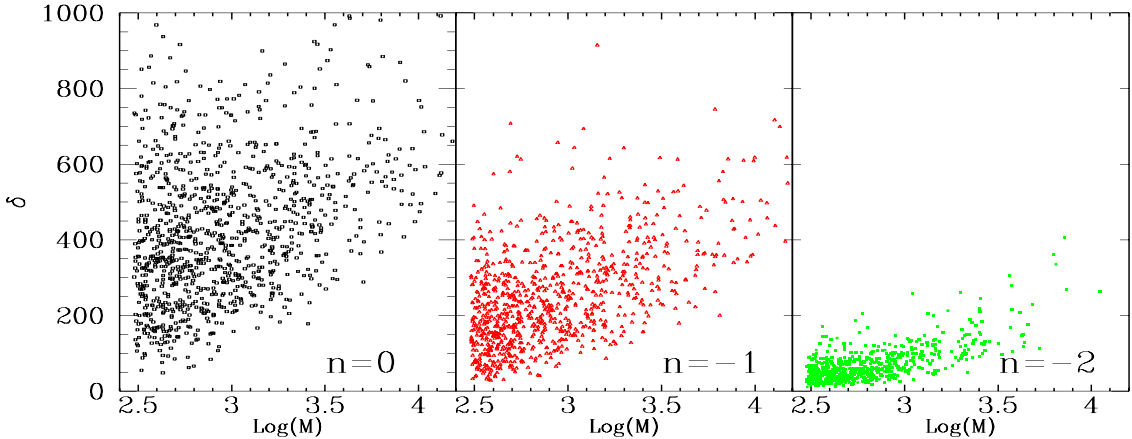,height=6.0cm,width=16cm}
\caption{Average density of structures found by DENMAX plotted against
  the logarithm of their masses for the  three different initial power
  spectrum.}
\label{denmax_densite}
\end{figure*}
For this algorithm the free parameter is the resolution of the grid on
which the density field is  computed.  Since we  use a PM code we have
chosen the same grid as the  one used during  the simulation.  Using a
finer grid  would not  make much sense  since density  maxima within a
cell of the  PM do not  have any meaning.   On Fig.  2 we present  the
multiplicity  function obtained with  DENMAX and rescaled  to $a=1$ as
before.   For $n=0$ the  time scaling property  is well recovered even
though one starts to see a  little shift in the  small mass region and
the  agreement with a PS  function  with $\delta_c=1.4$  is striking.  
However for $n=-1$   and $n=-2$ there is  a  clear shift  between  the
multiplicity function at  different times.  The  time scaling property
is not satisfied at all by the structures  recognize by DENMAX. As can
be seen  on Fig. 4-B where  we have plotted the universal multiplicity
function of DENMAX for the three different power spectrum the spectral
invariance is not  satisfied either (We have   done an average of  the
multiplicity function   at different times  in order  to have a single
curve even though  it is not legitimate in  this case).  This fact can
also  be seen on  Fig.  2 by looking  at  the relative position of the
multiplicity  function compared to the PS  multiplicity function.  The
large  modification   of the  DENMAX  multiplicity  function  when one
changes the spectrum   of  initial fluctuation  can be  understood  by
looking      more  closely at   the    structures   found.    In Fig.  
\ref{denmax_densite} we have  plotted  the average density  versus the
mass of all the structures found for the different spectra.  It is now
quite clear that the structures found by  DENMAX greatly depend on the
spectra.  For $n=0$  the density wells  are quite steep  and therefore
the DENMAX  structures are dense.     On the contrary for  $n=-2$  the
density  wells are rather flat which  leads to less  dense structures. 
More  generally, DENMAX detects as   structures regions which are  not
necessarily  dense.  In  order   to  correct  this drawback   we  have
developed the algorithm DENEVAP presented above.  In Fig. \ref{fmlog1}
the multiplicity function  obtained   with DENEVAP is presented.    By
taking away the  low density region we have  improved the time scaling
at high  mass, but it is still  very bad  (worse) for intermediate and
low mass.  It is of course  possible to truncate the DENMAX structures
in many  other  ways but the results  should  not  be  very different. 
Beyond the problem of the  structures' densities, the DENMAX algorithm
is not very well suited to the detection of cosmic structures because,
with  the  grid used  to  compute the density   field, it introduces a
characteristic  scale,  $L$.  This  length, fixed   once and for  all,
correspond to a  density  fluctuation,  $\Delta$, which  is  different
depending on the time and on the spectrum considered. For this reason,
the two scaling behaviors are broken.  To recover them, the length $L$
could be determined as a function  of the time  and the power spectrum
in  order to correspond  to a constant  $\Delta$.  But this artificial
introduction  of  the scaling laws   in the  algorithm used to  detect
structures does not seem satisfactory to us.
 
\subsection{Spherical Overdensity}

The only free parameter of this algorithm is the density threshold. To
compare with FOF  we have tested  two  thresholds: 180  (SO180) and 60
(SO60) which  roughly correspond to FOF  with a percolation  length of
$0.2$ and $0.3$  respectively.   The resulting multiplicity  functions
are presented on Figs. \ref{fmlog1} and \ref{fmlog2}.  The time self
similarity is roughly  reproduced for SO60 but  very badly for  SO180. 
We have therefore also  looked at the  spectral  invariance for  SO60. 
The results are much better than those of DENMAX but there is a slight
shift.    FOF  reproduces this  invariance   much better.   For $n=-2$
structures are rather isolated from one another and therefore imposing
a spherical shape is not a strong  constraint.  However for $n=-1$ and
$n=0$ structures are more strongly correlated and imposing a spherical
shape on  an heavy cluster damages  close-by  structures.  SO180 works
poorly because  imposing a spherical shape  to  structures with such a
high density criteria results in the truncation of some dense parts of
elongated structures.  For    SO60  this truncation  occurs  less  and
concerns region of inferior density contrast.

\begin{center}
\begin{figure*}[httb]
\psfig{file=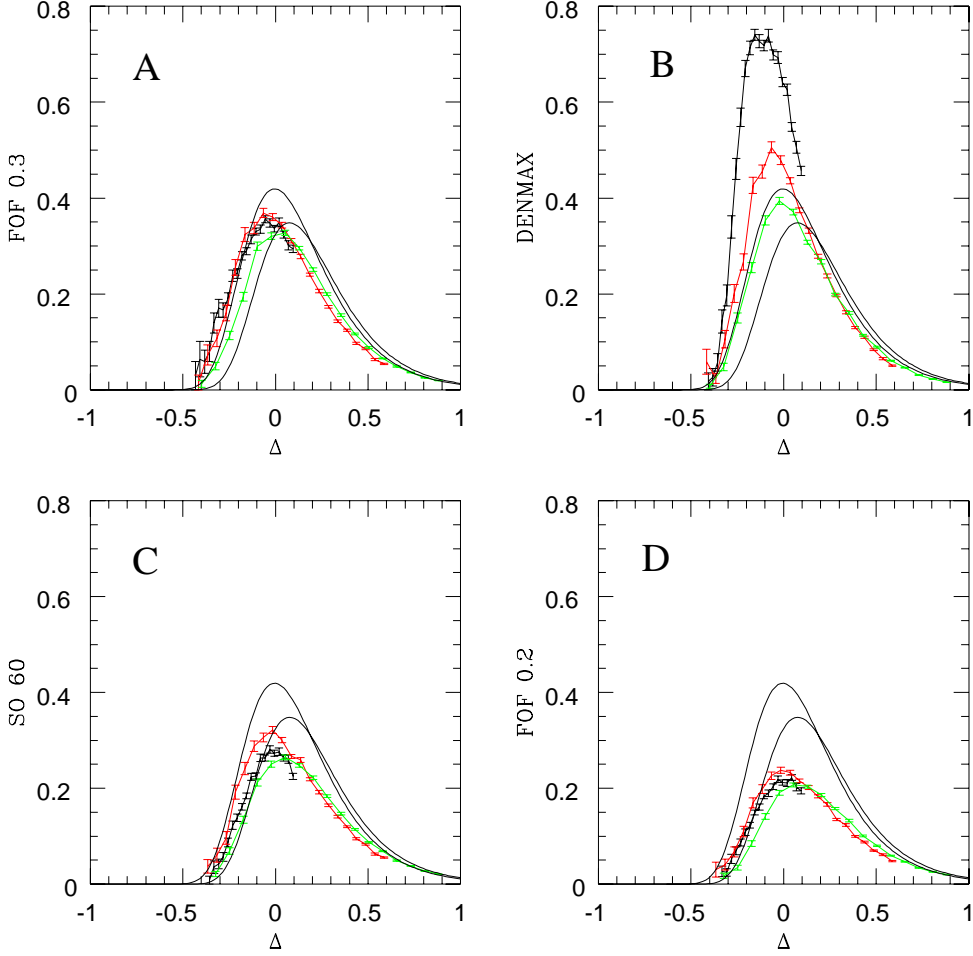,height=14cm}
\caption{Universal multiplicity functions obtained for FOF-0.3,
  DENMAX,  SO60  and FOF-0.2.  Each  curve  corresponds to a different
  power spectrum.  As before, the  solid  lines correspond to PS  mass
  function with $\delta_c=1.4$ and $1.686$.}
\label{fmuni}
\end{figure*}
\end{center}

\section{Link with the initial conditions}

In  the previous sections we  have analyzed  the agreement between the
multiplicity function of numerical structures  recognized with a given
algorithm  and the ones obtained   in  a PS  formalism  using a  given
dynamics.  This analysis was only statistical and its sole concern was
the number of structures. However, this point of view is not enough to
have a  coherent description of  structure  formation.  If a numerical
and an analytical multiplicity function  are identical it is then also
necessary that the numerical  structures  follow the dynamics used  in
the analytical picture.  Otherwise  the agreement of  the multiplicity
functions is  a fit rather than  a  dynamical description of structure
formation. In the PS formalism the dynamics intervene only through the
collapse time. We have therefore decided  to compare the collapse time
of the structure predicted by dynamical models and the one obtained in
the  simulation.  In   order to  give  a  satisfactory  description of
structure formation, the PS formalism should  agree with the numerical
simulations  both in the initial and  final conditions and also in the
dynamical link between these two states. We will test the agreement in
the initial conditions by comparing the  initial distribution of shear
and density contrast of the  selected halos, for the final  conditions
we will use the multiplicity function and the dynamical coherence will
be tested by comparing the collapse time.

We  have  done this global  analysis  with the  halos detected  by the
friend-of-friend because they satisfy better the two scaling criteria.

\subsection{Computation of the initial parameters of halos}

The first step of this analysis is to determine the initial properties
of the halos detected in the final  state of the numerical simulation. 
Once a group of particles has  been identified as  a halo of the final
density field, we  determine its initial lagrangian  volume (i.e.  the
volume  occupied by theses particles  in  the initial conditions).  We
then  compute the density field in  the initial condition with a cloud
in cell scheme  using all the particles of  the simulation.  The shear
which is  the  traceless  symmetric  part  of the  tensor   $(\partial
v_i/\partial x_j)$, is then deduced  from the density field using  the
relations $\partial v_i/\partial x_j  = \partial^2 \delta/\partial x_i
\partial x_j$  valid in the  linear regime.   The initial  density and
shear of a halo are then defined as the average of the previous fields
over its  lagrangian  volume.   The  resulting shear tensor   is  then
diagonalized to obtain the initial shear eigenvalues.

With  the previous  procedure,  the density  contrast   that we obtain
corresponds to the value of the  initial density field smoothed with a
top-hat   filter having the  shape of  the  initial lagrangian volume. 
However, the variable used in the PS formalism  is the initial density
field smoothed by  a {\bf spherical}  top-hat filter.  It is therefore
necessary to check that    these  two density fields have    identical
statistical properties.  This has been done  by comparing directly the
probability distribution  function (PDF) of  both fields  (for several
halos  shapes).  The  maximum discrepancy between  the   two PDF's was
found to  be in all  cases less  than $5\%$.   If  these two smoothing
methods are statistically equivalent,  they however differ drastically
on a halo-by-halo basis. The initial  conditions attributed to a given
structure will be very different depending on whether they are computed
in a  sphere or in the initial  lagrangian volume.  From an analytical
point of view, the choice of the filter  has no influence on the shape
of the mass function for scale-free initial conditions (\cite{LC93}).

\subsection{Toward a coherent picture ?}

Now  that  we  have both   the initial properties  of  the  structures
detected by FOF-0.2 and  FOF-0.3 and their multiplicity functions,  we
are in a position to test the dynamical meaning of the PS formalism as
a tool to study structure formation.

Analytically, structures can be identified, as we have stressed in P1,
with regions which have collapsed either  along their first, second or
third principal axis.  Before  looking for a precise  dynamical model,
it is  therefore necessary to determine which  of these  three type of
collapses  corresponds  to  the    structures   found  in    numerical
simulations.  As we have shown in P1, the  derivative of the selection
function makes a direct link between the multiplicity function and the
underlying   dynamical model.  For  this  reason, it has very specific
behaviors corresponding  to each of   these  three type of  dynamics.  
Therefore, we  will   find if  the structures found   in the numerical
simulations  result  from a first, second   or third axis  collapse by
looking at the derivative of their selection function.

\begin{figure*}
\psfig{file=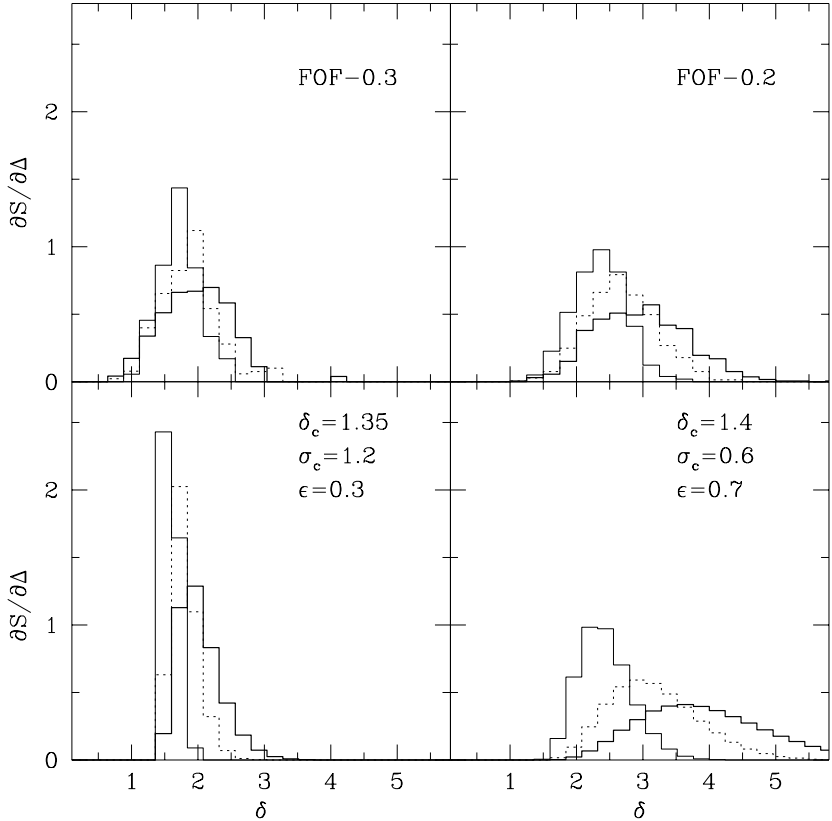,height=12.0cm,width=12.0cm}
\caption{Derivative of the selection function obtained from numerical 
  simulation data for FOF-0.3 (upper-left panel), FOF-0.2 (upper-right
  panel). The two lower panels present the same function corresponding
  to  the  analytical  model  (see  text  for the   definition  of the
  parameters) using  the same bin  size.  The solid,  dotted and thick
  lines correspond respectively to $\Delta=$1.5, 2.5 and 3.5. }
\label{fonc_sel_num}
\end{figure*} 

From formula \ref{phi_uni} (in the appendix), it is possible to deduce
that the  function $P_{\Delta}(\delta) \propto e^{-\delta^2/2\Delta^2}
\partial S(\delta,\Delta)/   \partial \Delta  $ gives the  probability
that a halo of mass $\Delta$ has an initial density contrast $\delta$.
Choosing a mass range   between $\Delta$ and $\Delta+d\Delta$,   it is
possible to compute   this initial probability   distribution function
using the results  of our numerical  simulations and then to infer the
derivative  of   the   selection function.     We   plot  on    Fig
\ref{fonc_sel_num} the derivative  of the selection function for three
different values of $\Delta$.  The observed behavior (the position and
the width of the   peak  are  increasing  functions of   $\Delta$)  is
characteristic of a third axis collapse (P1).   The collapse along the
third axis  is slowed down by the  effect of the shear.  Therefore, on
small scales (high $\Delta$) where the shear is statistically greater,
structures need on average a higher  density contrast to collapse.  We
therefore deduce qualitatively that the underlying dynamical model, in
the framework of the PS formalism, is better described by the collapse
of the third axis.

Since the collapse   of the first axis is   a singularity, it   is not
possible follow the dynamics  of the structure  until the collapse  of
the   third  axis. However,   extrapolating   results obtained for the
collapse time of the first axis, we have  proposed in P1 the following
dynamical prescription which allows  to compute the collapse epoch  of
the third axis, as a function of the initial density contrast $\delta$
and the largest (and therefore positive) shear eigenvalue $\sigma$

\begin{equation} 
\frac{1}{a_c}          =            \frac{\delta}{\delta_c}
\left(1-\frac{\sigma}{\delta\sigma_c}\right)^{\epsilon}
\label{3param}
\end{equation} 

\noindent
where  $\delta_c$  is a density  threshold,  $\sigma_c$ is the maximum
allowed  initial  shear-to-density ratio   beyond which  shear effects
prevent  any halo  from  collapsing.  The last  parameter, $\epsilon$,
determines  the global shape of the  collapse  epoch.  From asymptotic
arguments, we stressed   in P1 that $\epsilon$   has to  be less  than
unity.

In  order  to recover a  coherent  description  of structure formation
within  the PS approach,  we   have verified that    a single set   of
parameters is able   to   reproduce simultaneously the    multiplicity
function (final state),  the  mean and  the  variance of the  function
$\partial S/\partial \Delta$ (initial state), and finally the collapse
time of halos.

\begin{figure}[httb]
\psfig{file=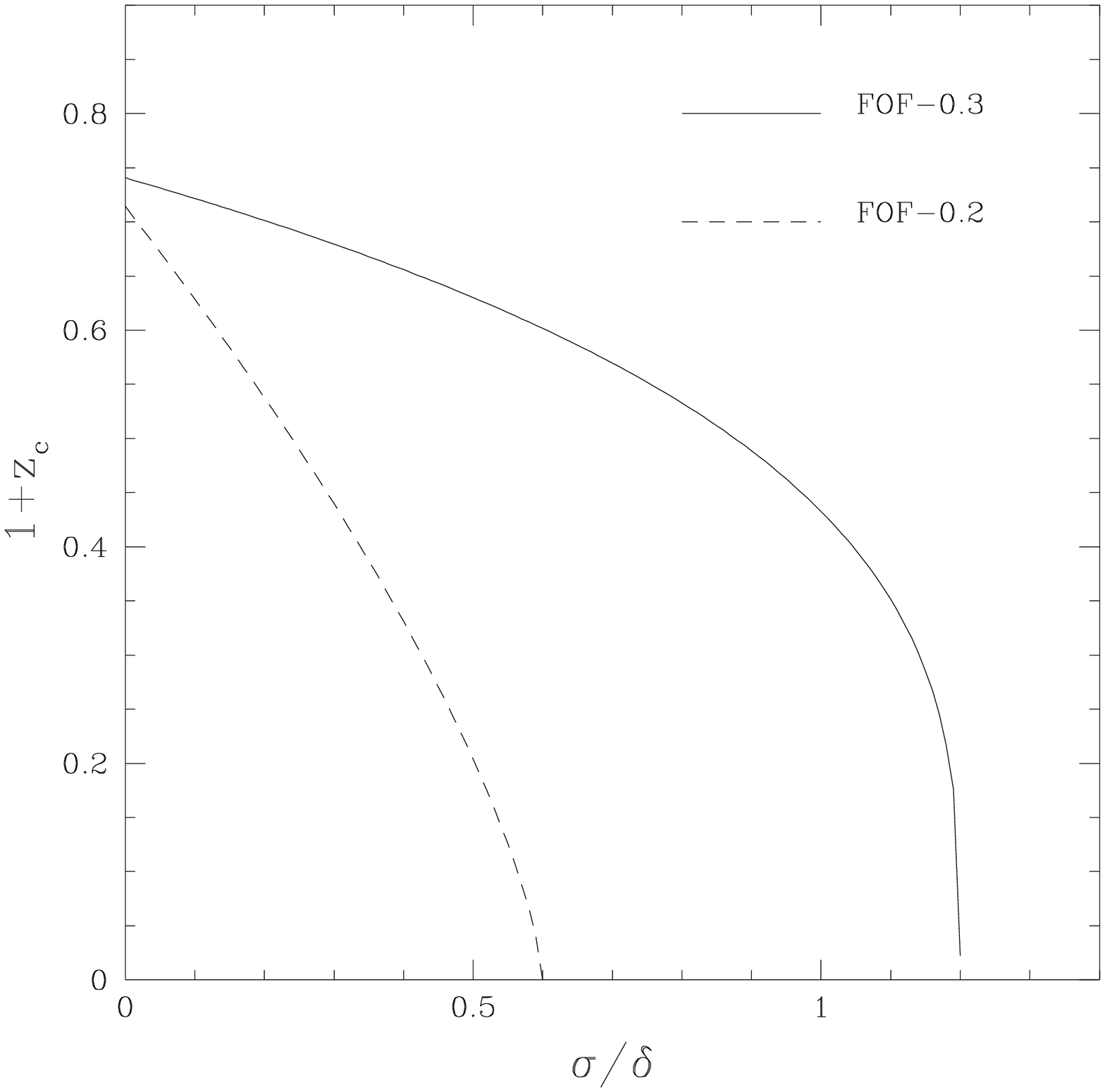,height=9cm}
\caption{Inverse of the collapse time (formula (\ref{3param}) for
  $\delta=1$ as a function of the shear. The solid (resp. dashed) line
  represents the model corresponding to FOF-0.3 (resp. FOF-0.2).}
\label{model_tc}
\end{figure}

\begin{figure*}[httb]
\psfig{file=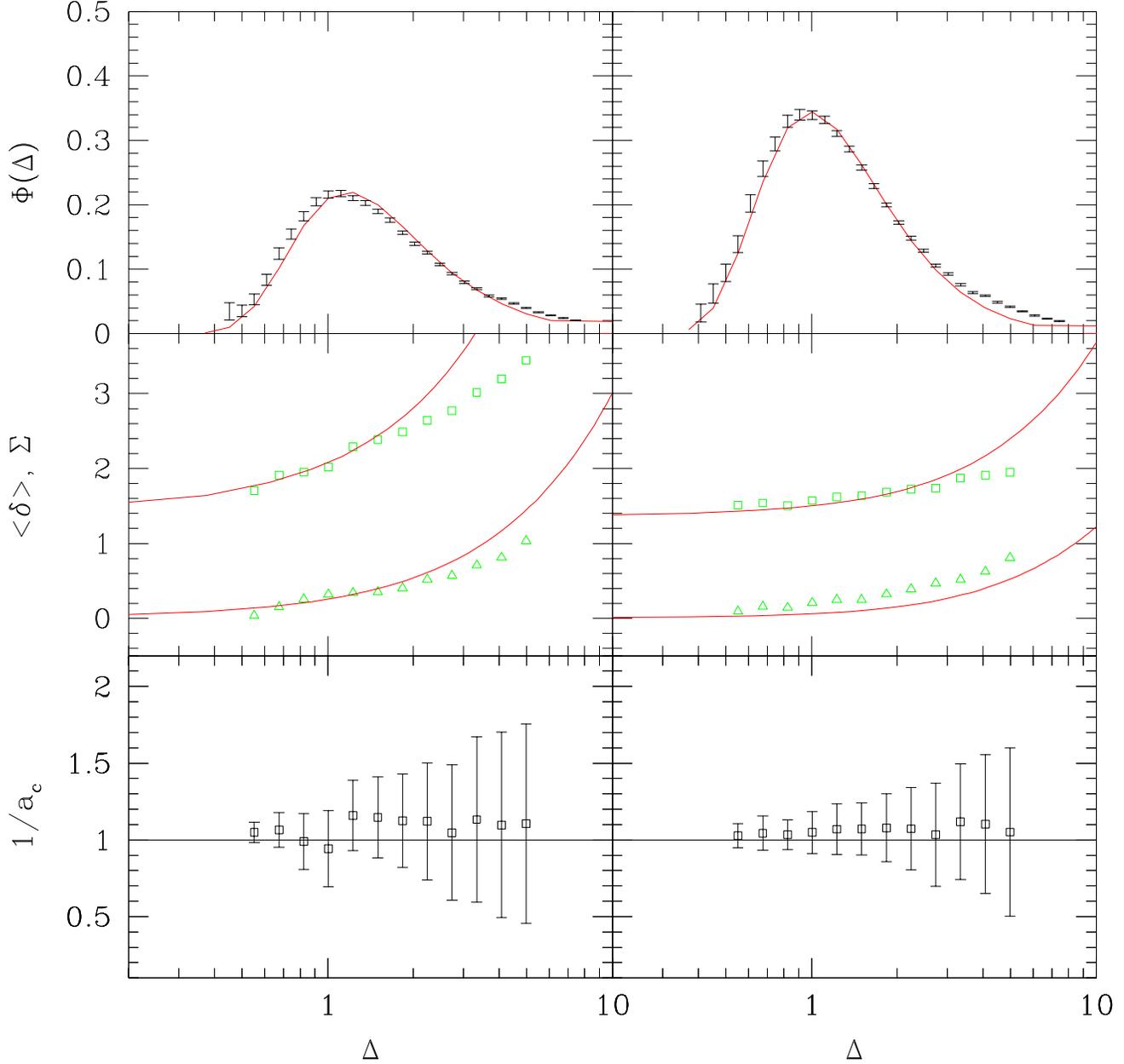,height=18cm}
\caption{The two upper panels show the numerical mass function obtained for
  FOF-0.2 (left)  and FOF-0.3 (right). The  solid curves are  the mass
  function obtained  with the analytical  model presented in the text. 
  The   middle   panels  show   the average   (square)  and dispersion
  (triangle) of the   function $\partial S/\partial \Delta$  with  the
  curve corresponding to the model.  The lower panels shows the average
  and  the  dispersion of the  collapse time  found  for the numerical
  structures with the analytical model.}
\label{global_fit}
\end{figure*}

For FOF-0.3 (resp. FOF-0.2) we find  that a good agreement is obtained
with $\delta_c=1.35$,    $\sigma_c=1.2$  and   $\epsilon=0.3$   (resp. 
$\delta_c=1.4$, $\sigma_c=0.6$ and $\epsilon=0.7$). The  corresponding
collapse epoch are plotted in Fig. \ref{model_tc}.  The two collapse
times corresponding  to each algorithm are  quite different.  But they
however result  from  the   same underlying   non-linear gravitational
dynamics.  The difference comes from  the fact that the two algorithms
define different   regions in the  final  density  field  and that the
initial physical conditions necessary to the  formation of such region
are thus  different.  The  collapse time  corresponding  to FOF-0.2 is
always larger.  This  is natural since it  takes more time for a given
halo to reach a higher density  contrast.  The critical shear value is
greater  for FOF-0.3 than for FOF-0.2.   This  was also to be expected
since, in the  context of our dynamical  model, the shear always slows
down   the collapse.  The   largest peaks of  the  final density field
therefore come from region  with a low  shear to density  ratio.  This
result is   consistent   with   those   obtained by   other    authors
(\cite{vdW94}, \cite{Ber94}), but we analyze  it from a very different
point of view.  The fact that  the dense regions  in the final density
field have  less  initial shear does  not mean  that  the dynamics  is
quasi-spherical, but, on the contrary, that the dynamical influence of
the shear is  such that dense structure  can form only in region where
it is sufficiently low. The dynamics of the densest structures is {\bf
  a   posteriori} spherical  (because  the spherical   collapse is the
fastest) but  one cannot suppose {\bf a  priori} that  the dynamics is
spherical everywhere.

We plot in  Fig. \ref{global_fit}  the  best agreement  obtained for
FOF-0.2  and   FOF-0.3 with  the  collapse    time given  by  equation
(\ref{3param}) and the above  parameters.  We reproduce in both  cases
very    accurately the    multiplicity    function,  with   a   slight
underestimation at the very low  mass end.  Note  that the error  bars
are only Poisson  noise estimators which  do not take into account the
systematic errors coming  from the finite  resolution  of the PM  code
such as, for example,  the loss of  dynamical accuracy in the halos of
lower  mass.  Moreover, we also  reproduce the mass  dependency of the
mean  ($<\delta>$) and  of   the variance  ($\Sigma$)  of the  initial
density contrast distribution of  the halos (Fig.   \ref{global_fit}). 
We  have also   plotted   in  Fig.  \ref{fonc_sel_num} the    entire
derivative of the selection function.

Finally,  the  mean  collapse  epoch is  recovered  ($a_c \simeq  1$),
although the  dispersion remains large for  small mass  objects.  Note
that  for small mass  halos, in addition  to lower dynamical accuracy,
there is also a  systematic error  in  the computation of  the initial
parameters.  In  fact,  due to the   discreteness of the computational
grid,  the error done when  averaging over  the Lagrangian volume then
increases because boundary effects become important.

\begin{center}
\begin{figure*}[httb]
\psfig{file=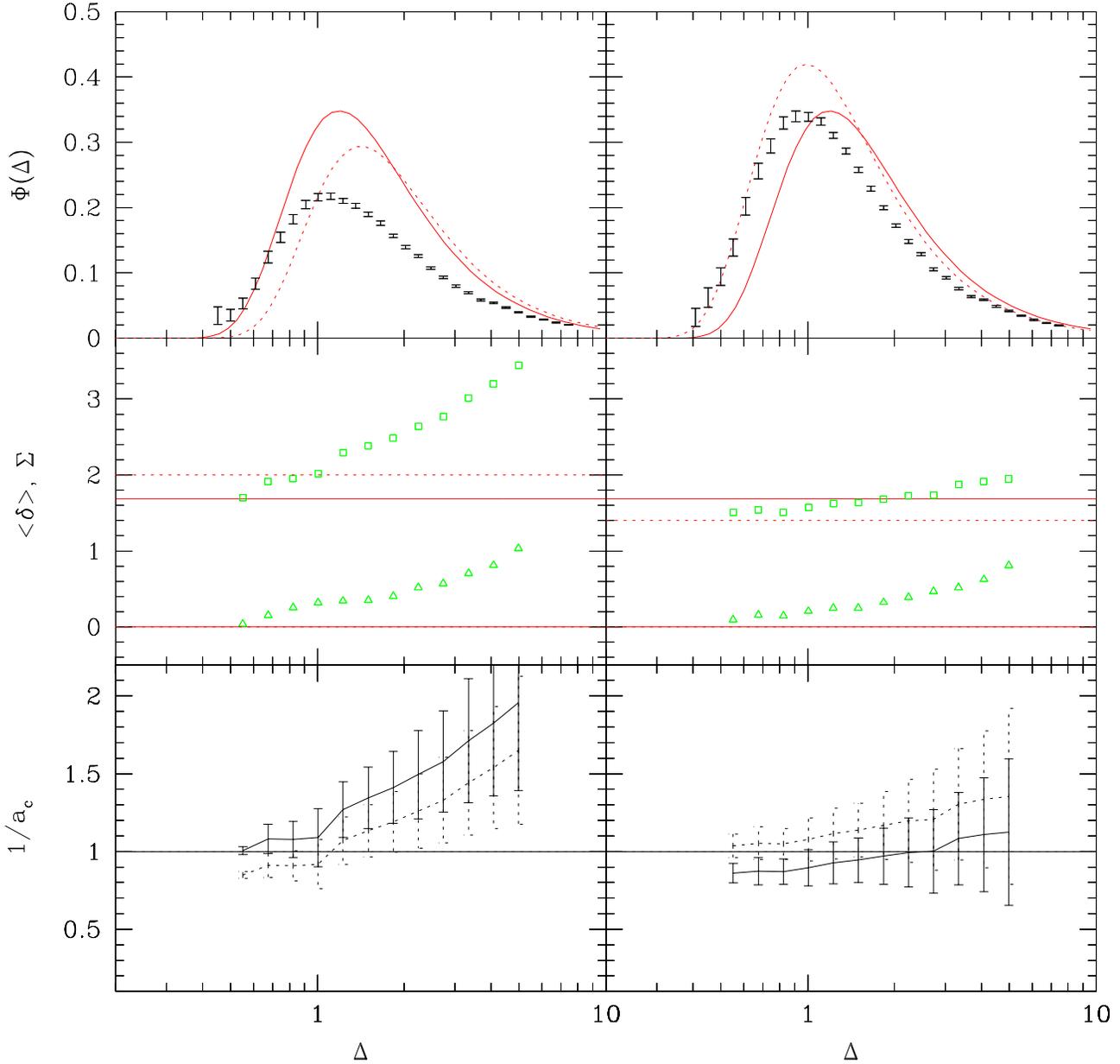,height=18cm}
\caption{Same as figure (\ref{global_fit}) but for the classical
  PS formalism.  The solid lines  correspond to $\delta_c=1.686$.  The
  dashed   lines correspond   to    $\delta_c=2.$ (left   column)  and
  $\delta_c=1.4$ (right column).}
\label{pscomp}
\end{figure*}
\end{center}

These results show altogether that our  formalism is able to reproduce
the mass  function both from a  statistical and from a dynamical point
of view.  However, each halo does  not individually follow exactly our
simple dynamical prescription but in each mass range the collapse time
of  the  halos is on   average given by our  model.    For the sake of
comparison,  we plot in Fig.  \ref{pscomp} the same results obtained
with the  standard PS approach,   using  $\delta_c$ as the  only  free
parameter.  We  plot  in each  panel  the  results obtained using  two
different  values  for $\delta_c$, showing  that   it is impossible to
reproduce  simultaneously  the final   and initial  states.   In other
words, the standard PS approach gives a rough statistical indicator of
the number of halos, but does not allow for  a coherent description of
structure formation.

\section{Conclusion}

In P1 we have extended the PS formalism in order to include the effect
of the shear and the tide. We have shown  than within the PS formalism
it was   possible to  obtain  very different  behaviors for   the mass
function which are directly related to the underlying dynamical model.
Even though the mass functions are quantitatively very different, they
all satisfy, within the PS formalism, two scaling properties (i.e. the
time   self similarity and     the  power spectrum  scaling   (formula
(\ref{spec-scaling})). The  aim of this  paper was to determine if the
PS formalism could explain coherently  structure formation both from a
statistical and a dynamical point of view.

In  the  first part  of  this  paper we  have  shown,  using numerical
simulation, that it was  possible to define numerical structures whose
mass function obeys both the PS scaling laws.  We have tried different
algorithms which all  lead to different mass  functions.  The only one
which reproduces all the scaling laws is friend--of--friend.  However,
the resulting universal mass function  is still strongly dependent  on
the  percolation length.  This dependence  of the mass function on the
chosen algorithm shows that the choice of the dynamics  used in the PS
formalism  cannot  be unique  and  has to  be  related to  the  actual
definition of the structures in the density field.  In order to have a
coherent description of    structure formation, the  chosen  dynamical
model should not  only reproduce the  mass  function, but  should also
give  a correct description  of the  regions selected  in the  initial
density field  and of their  collapse  times.  We have  shown, using a
dynamical  prescription proposed in P1,  that it is  possible to built
such  a coherent picture.  Within  this picture, the structures result
from a collapse along  their third principal  axis. The collapse epoch
is then  a function of the density  contrast and  of the largest shear
eigen-value  given by  formula (\ref{3param})   which has three   free
dynamical parameters.  It is then possible to find a set of parameters
which    reproduces simultaneously the    mass  function, the  initial
selection function  and gives on average  the correct  collapse epoch. 
Such a  coherent description is  totally impossible with the classical
PS    formalism  even when   adjusting  the   density threshold.  This
inadequacy comes from the fact that  the shear plays an important role
in the dynamics of forming  halos.  This dynamical effect explains the
$50\%$  discrepancy of  structures  found by  FOF-0.2  compared to the
classical PS formalism.

\appendix
\section{Appendix}
The {\bf  mass function}, $\Phi(M)$ which gives  the number density of
object  of mass $M$, and   the {\bf multiplicity  function} $\mu (M)$,
which gives the fraction of mass embedded in objects of masses between
$M$ and $M+dM$ are simply related by: $\Phi(M)=\frac{\rho_0}{M}\mu(M)$
where $\rho_0$ is  the mean density of the  universe. In the framework
of a PS approach the multiplicity function can be expressed as:

\begin{equation}
\label{spec-scaling}
\mu(M)=\frac{d\Delta}{dM}\mu_{u}(\Delta) 
\end{equation} 

\noindent where $\Delta$ is the r.m.s density contrast filtered by a
Top -- Hat window function of  radius $R$ (This corresponds to objects
of mass $M  = (4\pi/3) \rho _0 R^3$).  The first term depends  only on
the  power spectrum  of the   initial density  fluctuation, while  the
second  term which we call  the  {\bf universal multiplicity function}
depends only on the dynamical prescription describing the formation of
structure.  Assuming spherical dynamics $\mu$ takes the form:
\begin{equation}
\mu (M) = -\sqrt {\frac{2}{\pi}}\frac{\delta_c}{\Delta
  ^2}\frac{\partial \Delta}{\partial
  M}\hbox{exp}\left(-\frac{\delta_c^2}{2\Delta^2} \right)
\end{equation}
\noindent with $\delta_c=1.686$.   More generally, the universal
multiplicity function can be expressed as (P1)

\begin{equation} 
\label{phi_uni}
\mu_{u} (\Delta)=- C
\int _{-\infty} ^{+\infty}
e^{-\nu^{2}/2} \frac {\partial S}{\partial \Delta} d\nu
\end{equation}
where $C$ is a normalization constant. The derivative of the selection
function, $S$,  depends only on  the dynamical model chosen to compute
the collapse time of structures.  All the dynamical information needed
in the PS formalism is contained in this function.


\begin{thebibliography}{}
\bibitem[Audit et al. 1997]{AA97} Audit E., Teyssier R., Alimi
  J-M.,  1997, A\&A, {\bf 325}, 439-449

\bibitem[Bertschinger \& Gelb 1991]{BG91} Bertschinger E., 
 Gelb J., 1991, Comp. in Phys., {\bf 5}, 164

\bibitem[Bernardeau 1994]{Ber94}Bernardeau F., 1994, ApJ, {\bf 427}, 51

\bibitem[Efstatiou et al. 1985]{EDWF85}Efstathiou G.,
  Davis M.,  White S.D.M., Frenk C.S., 1985, ApJS, {\bf 57}, 241  

\bibitem[Efstatiou et al. 1988]{EFWD88}Efstathiou G.,
  Frenk C.S., White S.D.M., Davis M., 1988, MNRAS, {\bf 235}, 715  

\bibitem[Eke et al. 1996a]{Eke96a} Eke  V.,
  Cole S., Frenk C., Navarro J., 1996, MNRAS, {\bf 281}, 703

\bibitem[Gelb \& Bertschinger 1994]{GB94} Gelb J., Bertschinger
  E., 1994, ApJ, {\bf 436}, 467
  
\bibitem[Lacey \& Cole 1993]{LC93}Lacey C., Cole S., 1993, MNRAS,
  {\bf 262}, 627
  
\bibitem[Lacey \& Cole 1994]{LC94}Lacey C., Cole S., 1994, MNRAS,
  {\bf 271}, 676

\bibitem[Press and Schechter 1974]{PS74} Press W.H., Schechter P., 1974,
  ApJ, {\bf 188}, 425

\bibitem[Van de Weygaert \& Babul 1994]{vdW94} Van de Weygaert R.,
  Babul A., 1994, ApJ, {\bf 425}, L59

\end{thebibliography}
\end{document}